# On Coparanormality in Distributed Supervisory Control of Discrete-Event Systems

Vahid Saeidi[1], Ali A. Afzalian[2] and Davood Gharavian[3]
Department of Electrical Eng., Abbaspour School of Engineering,
Shahid Beheshti University, Tehran, Iran
[1] v_saeidi@sbu.ac.ir, [2]Afzalian@sbu.ac.ir, [3]d_gharavian@sbu.ac.ir

## Abstract

Decomposition and localization of a supervisor both are reduction methods in distributed supervisory control of discrete-event systems.Decomposition is employed to reduce the number of events and localization is used to reduce the number of states of local controllers.In decomposition of a supervisor both observation and control scopes are restricted, whereas in localization only control authority is restricted to the corresponding local controller. In this paper, we propose a decomposition method by defining coparanormality property, and by using relative observability property of a monolithic supervisor. Coparanormality is a coobservation property defined based on paranormality property for a set of natural projections.It is shown that each supervisor can be coparanormal, provided a set of appropriate natural projections exist. Moreover, it is proved that relative observability is a sufficient condition for decomposition of a supervisor. Furthermore, the supervisor localization procedure is generalized to find a set of local controllers for any partition ofthe controllable events set. The implementation of such local controllers may become easier in industrial systems.

## 1. Introduction

In the supervisory control of discrete-event systems (DES), the monolithic (global) supervisor has enough information to satisfy the designed specification. Since the supervisory control synthesis faces to computational complexity, modular [1-3], hierarchical [4], and heterarchical [5, 6] synthesis methods and the nondeterministic automata approach [7, 8] have been practiced to handle the computational complexity.

Decentralized supervisory control has been proposed to reduce the computational complexity in large scale DES [9-11]. Since a decentralized supervisor has partial observation of the plant, it does not have enough information about other supervisors and may be in conflict with them. Also, there has not been proposed a guideline to find the significant events for making consistent decisions in each decentralized supervisor, so far.Decomposability and strong decomposability (conormality) were introduced in [9]to construct decentralized supervisory control in a top-down approach. Recently, some works have been carried out to find a decomposable sublanguage of a specification [12].

Since modular and decentralized supervisory controls are the methods with a bottom-up approach, construction of a coordinator was proposed to remove the conflict between decentralized supervisors [13-15]. Although the coordinator removes conflicts in decentralized supervisory control, it does not allow decentralized controllers to operate independently without conflict. Also, some research works have been reported on detecting conflict between decentralized supervisors using the observer property of natural projection [16].

Distributed supervisory control, constructed by supervisor localization procedure, guarantees no conflict between local controllersin a top-down approach. The goal of supervisor localization is to realize performance identical to that achieved by the monolithic control. It is also desired that each localized controller be as simple as possible, so that individual strategies are more comprehensible, among diverse criteria of simplicity, e.g. reducing the state size. The localization procedure is conducted based on control information directly relevant to the target agent; this way is carried out for each agent in the plant, individually. In large scale systems, owing to state space explosion, the monolithic supervisor might not be feasibly computable. Combining supervisor localization with the efficient modular control theory [13] was proposed in [17] to manage such complexity. However, the control authority of a local controller is strictly local; the observation scope of each local controller is systematically determined in order to guarantee the correct local control action. Moreover, synchronization of local controllers with the plant is control equivalent to the monolithic supervisor with respect to (w.r.t.) the plant [17].

On the other hand, partial observation-based supervisory control has been introduced to overcome lack of enough sensor and inadequate information of the plant for consistent decision making. Observation properties such as normality, observability [18, 19] and relative observability [20] describe the effect of observation on behavior of the supervisor. Paranormality is another property of a language, i.e., the occurrence of unobservable events in the plant never causes an exit from the closure of the language.

In this paper, we have two main goals: 1. Generalization of the supervisor localization procedure corresponding to any partitioning of the controllable events set, 2. Decomposition of a supervisor using relative observability of a monolithic supervisor and by defining coparanormality property. For this purpose, at first we show that the supervisor localization procedure can be assumed to be separated into two steps:

1. Consider a partition of the controllable event set corresponding to controllable events of each component,and make self-loop all eventsbelong to each subset at the states of the supervisor, where such events are disabledin the monolithic supervisor.In this step a set of local controllers is constructed.

2. Reduce the number of states in each local controller by supervisor reduction procedure to make a reduced state local controller (in the sense of [17]).

Next, we show that the supervisor localization procedure can be generalized to localize a monolithic supervisor w.r.t. anyarbitrary subset of controllable events. Since the proposed method is a top-down approach, we should have a synthesized monolithic supervisor.Meanwhile, the proposed method can be used to localize a coordinator andeach modularsupervisorw.r.t. any arbitrarysubset of corresponding controllable events in large scale systems.

We define a new observation property, i.e. coparanormality, in order to distribute the supervisory control of DES. Coparanormality is defined based on paranormality of a supervisor w.r.t. the closed language of the plant and a set of natural projections. It is shownthat each supervisor can be coparanormal by defining an appropriate set of natural projections. We prove thateach local controller, constructed based on coparanormality property, is control equivalenttothe corresponding local controller which can be constructed by the supervisor localization procedure w.r.t. the plant. Also, it is proved thatrelative observability property is a sufficient condition for decomposability of a supervisor, i.e., coparanormality property becomes decomposability in a relative observable supervisor. We illustrate in Section 6, the constructed local controller corresponding to each controllable event has less number of states rather than local controllers which are constructed corresponding to components of the plant. This method can be beneficial when each component has more than one controllable event. Also, the components of the plant need not to have disjointed events. The implementation of such local controllers may become easier in industrial applications.

The rest of the paper is organized as follows: In Section 2, the basic notions of supervisory control theory are reviewed. In Section 3, supervisor localization procedure is generalized for any subset of controllable events. In Section 4, coparanormality property is introduced and a method is proposed to distribute a supervisor based on this property. In Section 5, we prove that the relative observability of a supervisor is a sufficient condition for its decomposability. In Section 6, the extended theories are illustratedbythe supervisory control of guide way. Finally, concluding remarks are outlined in Section 7.

## 2. Preliminaries

A discrete-event system is presented by an automaton$\mathbf{G} = (Q, \Sigma, \delta, q_0, Q_m)$, where $Q$ is a finite set of states, with $q_0 \in Q$ as the initial state and $Q_m \subseteq Q$ being the marked states; $\Sigma$ is a finite set of events $(\sigma)$ which is partitioned as a set of controllable events $\Sigma_c$ and a set of uncontrollable events $\Sigma_u$. $\delta$ is a transition mapping $\delta: Q \times \Sigma \rightarrow Q$ , $\delta(q, \sigma) = q'$ gives the next state $q'$ is reached from $q$ by occurrence of $\sigma$. In this context $\delta(q_0, s)!$ means that $\delta$ is defined for $s$ at $q_0$.$L(\mathbf{G}) \coloneqq \{s \in \Sigma^* | \delta(q_0, s)!\}$ is the closed behavior of $\mathbf{G}$, and $L_m(\mathbf{G}) \coloneqq \{s \in L(\mathbf{G}) | \delta(q_0, s) \in Q_m\}$ is the marked behavior of $\mathbf{G}$ [5]. In the supervisory control context, a control pattern is $\gamma$, where $\Sigma_u \subseteq \gamma \subseteq \Sigma$ and the set of all control patterns is denoted with $\Gamma = \{\gamma \in Pwr(\Sigma) | \gamma \supseteq \Sigma_u\}$. A supervisor for a plant $\mathbf{G}$ is a map $V: L(\mathbf{G}) \rightarrow \Gamma$, where $V(s)$ represents the set of enabled events next to occurrence of string $s \in L(\mathbf{G})$. A pair $(\mathbf{G}, V)$ is written as $V/\mathbf{G}$. The closed loop language $L(V/\mathbf{G})$ is defined by: (1) $\epsilon \in L(V/\mathbf{G})(\epsilon \text{ is empty string})$, (2) $s\sigma \in L(V/\mathbf{G})$ iff $s \in L(V/\mathbf{G}), \sigma \in V(s)$, and $s\sigma \in L(\mathbf{G})$. The marked language of $V/\mathbf{G}$ is $L_m(V/\mathbf{G}) = L(V/\mathbf{G}) \cap L_m(\mathbf{G})$. The closed loop system is non-blocking if $\overline{L_m(V/\mathbf{G})} = L(V/\mathbf{G})$. $\overline{L_m(V/\mathbf{G})}$ is the set of all prefixes of traces in $L_m(V/\mathbf{G})$. A language $K \subseteq \Sigma^*$ is controllable w.r.t. $L(\mathbf{G})$ and uncontrollable events $\Sigma_u$, if $\overline{K}\Sigma_u \cap L(\mathbf{G}) \subseteq \overline{K}$. For every specification language $E$, there exists a supremal controllable sublanguage of $E$ w.r.t. $L(\mathbf{G})$ and $\Sigma_u$[15, 21].

A natural projection is a mapping $P: \Sigma^* \rightarrow \Sigma_0^*$, where (1) $P(\epsilon) := \epsilon$ , (2) for $s \in \Sigma^*$, $\sigma \in \Sigma$ , $P(s\sigma) := P(s)P(\sigma)$, and (3) $P(\sigma) := \sigma$ if $\sigma \in \Sigma_0$ and $P(\sigma) := \epsilon$ if $\sigma \notin \Sigma_0$. The natural projection $P$ can be extended and denoted with $P: Pwr(\Sigma^*) \rightarrow Pwr(\Sigma_0^*)$. The inverse image function of $P$ is denoted with $P^{-1}: Pwr(\Sigma_0^*) \rightarrow Pwr(\Sigma^*)$ for any $L \subseteq \Sigma_0^*$, $P^{-1}(L) \coloneqq \{s \in \Sigma^* | P(s) \in L\}$ [13]. The synchronous product of languages $L_1 \subseteq \Sigma_1^*$ and $L_2 \subseteq \Sigma_2^*$ is defined by $L_1 \parallel L_2 = P_1^{-1}(L_1) \cap P_2^{-1}(L_2) \subseteq \Sigma^*$, where $P_i: \Sigma^* \rightarrow \Sigma_i^*, i = 1,2$ for $\Sigma = \Sigma_1 \cup \Sigma_2$ [13]. A language$K \subseteq L(\mathbf{G})$ is decomposable w.r.t. $\mathbf{G}, P_1, P_2, \Sigma_c^1, \Sigma_c^2$, if $K = P_1^{-1}P_1(K) \cap P_2^{-1}P_2(K) \cap L(\mathbf{G})$. $K$is strongly decomposable (conormal) w.r.t.$\mathbf{G}, P_1, P_2, \Sigma_c^1, \Sigma_c^2$ if $K = (P_1^{-1}P_1(K) \cup P_2^{-1}P_2(K)) \cap L(\mathbf{G})$ [9]. These notions can be generalized for more than two projection channels. A language $K \subseteq \Sigma^*$ is $(L(\mathbf{G}), P)$-normal, if $P^{-1}P(\overline{K}) \cap L(\mathbf{G}) = \overline{K}$ [18]. $K$ is $(L(\mathbf{G}), P) -$paranormal if $\overline{K}(\Sigma - \Sigma_0) \cap L(\mathbf{G}) \subseteq \overline{K}$ [15]. If $\overline{K}$ is $(L(\mathbf{G}), P)$-normal, then it is $(L(\mathbf{G}), P) -$ paranormal. But the reverse is not true. Normality is a strong property and may not hold in practice. Another property has been defined, called relative observability. It imposes no constraint on disablement of unobservable controllable events, unlike the normality.

Consider $K \subseteq C \subseteq L_m(\mathbf{G})$, $K$ is relative observable w.r.t. $\bar{C}, \boldsymbol{G}$ and $P$, if for every pair of strings $s, s' \in \Sigma^*$ such that $P(s) = P(s')$, the following two conditions hold [20],

$$\begin{gathered}(i)(\forall\sigma \in \Sigma)\ s\sigma \in \bar{K}\ , s' \in \bar{C}, s'\sigma \in L(\mathbf{G}) \Longrightarrow s'\sigma \in \bar{K}, \\ (ii)\ s \in K, s' \in \bar{C} \cap L_m(\mathbf{G}) \Longrightarrow s' \in K.\end{gathered} \tag{1}$$

If $\bar{C} = \bar{K}$ then the relative observability definition becomes the observability definition. Let $\mathbf{SUP} = (X, \Sigma, \xi, x_0, X_m)$ be the recognizer for supervisor $K_s$, $\Sigma_0 \subseteq \Sigma$ and $P: \Sigma^* \to \Sigma_0^*$ be the natural projection. For $s \in \Sigma^*$, observation of $P(s)$ results in uncertainty as to the state of **SUP**given by the "uncertainty set" $U(s) \coloneqq \{\delta(q_0, s') | P(s') = P(s)\ , s \in \Sigma^*\} \subseteq Q$. Uncertainty set can be used to obtain a recognizer for $P(K_s)$. By definition of uncertainty set, each pair of states $x, x' \in X$, reachable by $s, s'$, are control consistent, if there exists a nonblocking supervisor $V$, such that $P(s') = P(s) \Longrightarrow V(s') = V(s)$. $V$ is called a feasible supervisor. A procedure was proposed in [15], to construct the projected DES $\mathbf{PS} = (Y, \Sigma_0, \eta, y_0, Y_m)$ and the corresponding feasible supervisor $\mathbf{SUP_f}$ to define the supervisory action of **PS** over the total events $\Sigma$.

The result of the procedure is $\mathbf{PS} = (Y, \Sigma_0, \eta, y_0, Y_m)$, where $Y$ is the final subset listing $\{y_0, y_1, \ldots\}$, $Y_m$ is the marked sublist such that $y \in Y_m$ iff $x \in y$ for some $x \in X_m$ and $\eta(y, \sigma) = y'$ iff $\xi(x, \sigma) = x'$ for some $x \in y$, $x' \in y'$ ($\sigma \in \Sigma_0$). In order to define the supervisory action of **PS** over the total events$\Sigma$, first introduce the disabling predicate $D(x, \sigma)$ to mean that $\sigma \in \Sigma_c$ and **SUP** disables $\sigma$ at $x$. Next introduce a partial function $F: Y \times \Sigma \longrightarrow \{0,1\}$ according to,

$$F(y, \sigma) = 0\ \text{ if } (\exists x \in y) D(x, \sigma). \tag{2}$$

It means that $\sigma$is controllable and is disabled at some $x \in y$.Also,

$$F(y, \sigma) = 1\ \text{ if} \sigma \in \Sigma - \Sigma_0 \& (\exists x \in y) \xi(x, \sigma)! \,\&\, [\ \sigma \in \Sigma_u \text{or } [\sigma \in \Sigma_c \& (\forall x' \in y) \neg D(x', \sigma)]]. \tag{3}$$

It means that $\sigma$ is unobservable and is enabled at some $x \in y$ and is either uncontrollable, or controllable and disabled in $y$. Otherwise, $F(y, \sigma)$ is undefined. Finally, modify the transition structure of **PS** to create $\mathbf{SUP_f}$ as follows,

$$\begin{gathered}(i)\ \text{If } F(y, \sigma) = 0, \text{delete any transition in } \mathbf{PS} \text{ of form } (y, \sigma, y'), \\ \text{i.e. declare} \eta(y, \sigma) \text{ undefined;} \\ (ii) \text{If } F(y, \sigma) = 1, \text{add the self} - \text{loop } \eta(y, \sigma) = y.\end{gathered} \tag{4}$$

The resulting structure $\mathbf{SUP_f}$ will be feasible and controllable. It is not guaranteed to be coreachable for a plant **G**, in general. If these properties happen to hold, then $\mathbf{SUP_f}$ provides a solution to the problem of feasible supervisory control.In this paper, the plant

**G** is assumed to be non-blocking, in order to guarantee coreachability of the feasible supervisor.

## 3. Supervisor localization versus supervisor reduction

A procedure was proposed in [22], to reduce the state size of a supervisor and it was generalized in [17], for supervisor localization.

Let $\mathbf{SUP} = (X, \Sigma, \xi, x_0, X_m)$ and define $E: X \to Pwr(\Sigma)$ as $E(x) = \{\sigma \in \Sigma | \xi(x, \sigma)!\}$. $E(x)$ denotes the set of events enabled at state $x$. Next, define $D: X \to Pwr(\Sigma)$as $D(x) = \{\sigma \in \Sigma \,|\, \neg\xi(x, \sigma)!\, \&(\exists s \in \Sigma^*)[\xi(x_0, s) = x \& \delta(q_0, s\sigma)!]\}$. $D(x)$ is the set of events which are disabled at state $x$. Define $M: X \to \{1,0\}$ according to $M(x) = 1$ iff$x \in X_m$, namely the flag of $M$ determines whether a state is marked in **SUP**. Also, define $T: X \to \{1,0\}$ according to $T(x) = 1$ iff $(\exists s \in \Sigma^*)\xi(x_0, s) = x \;\&\; \delta(q_0, s) \in Q_m$, namely the flag of $T$ determines whether some corresponding state is marked in **G**. Let $\mathcal{R} \subseteq X \times X$ be the binary relation such that for $x, x^{'} \in X$, $(x, x^{'}) \in \mathcal{R}$. $x$ and $x^{'}$are called control consistent, if

$$E(x) \cap D(x^{'}) = E(x^{'}) \cap D(x) = \emptyset, \tag{5}$$

$$T(x) = T(x^{'}) \Rightarrow M(x) = M(x^{'}). \tag{6}$$

Informally, a pair of $(x, x^{'})$ is in $\mathcal{R}$ if, by $(5)$, there is no event enabled at $x$ but disabled at $x^{'}$, and by $(6)$,$(x, x^{'})$ are both marked (unmarked) in **SUP**, provided that they are both marked (unmarked) in **G**. A cover $\mathcal{C} = \{X_i \subseteq X | i \in I\}$ of $X$ is called a control cover on **SUP** if [22],

$$(\forall i \in I) X_i \neq \emptyset \wedge (\forall x, x^{'} \in X_i)(x, x^{'}) \in \mathcal{R}, \tag{7}$$

$$(\forall i \in I)(\forall \sigma \in \Sigma)(\exists j \in I)\big[(\forall x \in X_i)\xi(x, \sigma)! \Longrightarrow \xi(x, \sigma) \in X_j\big], \tag{8}$$

Where, $I$ is an index set.

A control cover $\mathcal{C}$ lumps states of **SUP** into cells $X_i (i \in I)$ if they are control consistent. A control cover $\mathcal{C}$ is control congruence, if $X_i$ are pairwise disjoint. An induced supervisor is constructed as $\mathbf{J} = (I, \Sigma, \kappa, i_0, I_m)$ where $i_0 =$ some $i \in I$ with $x_0 \in X_i$, $I_m = \{i \in I | X_i \cap X_m \neq \emptyset\}$ and $\kappa: I \times \Sigma \to I$ with $\kappa(i, \sigma) = j$ provided, for such choice of $j \in I$,

$$(\exists x \in X_i)\xi(x, \sigma) \in X_j \& (\forall x' \in X_i)\big[\xi(x', \sigma)! \Rightarrow \xi(x^{'}, \sigma) \in X_j\big]. \tag{9}$$

A DES $\mathbf{RSUP} = (Z, \Sigma, \zeta, z_0, Z_m)$ is normal w.r.t **SUP** if,

$$\begin{gathered}(i)(\forall z \in Z)\big(\exists s \in L(\mathbf{SUP})\big)\zeta(z_0, s) = z,\\ (ii)(\forall z \in Z)(\forall \sigma \in \Sigma)[\zeta(z, \sigma)! \Rightarrow \big(\exists s \in L(\mathbf{SUP})\big)[s\sigma \in L(\mathbf{SUP}) \& \,\zeta(z_0, s) = z]],\\ (iii)(\forall z \in Z_m)\big(\exists s \in L_m(\mathbf{SUP})\big)\zeta(z_0, s) = z.\end{gathered} \tag{10}$$

As it was stated in [22], **RSUP** and **J** are DES-isomorphic. It was proved, **RSUP** and **SUP** are control equivalent w.r.t. **G**, i.e.

$$L_m(\mathbf{G}) \cap L_m(\mathbf{RSUP}) = L_m(\mathbf{SUP}), \tag{11}$$
$$L(\mathbf{G}) \cap L(\mathbf{RSUP}) = L(\mathbf{SUP}). \tag{12}$$

In [17], the authors assumed that **G** consists of component agents $\mathbf{G^k}$ defined on pairwise disjoint events sets $\Sigma^k$ ($k \in \mathcal{K}, \mathcal{K}$is anindexset). Let $L_k \coloneqq L(\mathbf{G^k})$ and $L_{m,k} \coloneqq L_m(\mathbf{G^k})$, the closed and marked languages of **G** are$L(\mathbf{G}) =\| \{L_k | k \in \mathcal{K}\}$ and $L_m(\mathbf{G}) =\| \{L_{m,k} | k \in \mathcal{K}\}$, respectively. Also,$\forall k \in \mathcal{K}$, $\bar{L}_{m,k} = L_k$ does hold, and **G** is necessarily nonblocking. With $\Sigma = \Sigma_c \dot{\cup} \Sigma_u$ a control structure is assigned to each agent $\Sigma_c^k = \Sigma^k \cap \Sigma_c$ ,$\Sigma_u^k = \Sigma^k \cap \Sigma_u$. A generator $\mathbf{LOC^k}$ over $\Sigma$ is a local controller for agent $\mathbf{G^k}$, if $\mathbf{LOC^k}$ can disable only events in $\Sigma_c^k$. Precisely, for all $s \in \Sigma^*$ and $\sigma \in \Sigma$ there holds,

$$s\sigma \in L(\mathbf{G}) \& s \in L(\mathbf{LOC^k}) \& s\sigma \notin L(\mathbf{LOC^k}) \Rightarrow \sigma \in \Sigma_c^k. \tag{13}$$

A set of local controllers $\mathbf{LOC} = \{\mathbf{LOC^k} | k \in \mathcal{K}\}$ is constructed, each one for an agent, with $L(\mathbf{LOC}) = \cap \{L(\mathbf{LOC^k}) | k \in \mathcal{K}\}$ and $L_m(\mathbf{LOC}) = \cap \{L_m(\mathbf{LOC^k}) | k \in \mathcal{K}\}$ such that **LOC** is control equivalent to **SUP** w.r.t. **G**.

In order to compute the supervisor localization procedure, define $E, M$ and $T$ same as the ones defined in supervisor reduction procedure. Next, define $D^k: X \to Pwr(\Sigma_c^k)$ as $D^k(x) = \{\sigma \in \Sigma_c^k \mid \neg\xi(x,\sigma)! \& (\exists s \in \Sigma^*)[\xi(x_0, s) = x \;\&\; \delta(q_0, s\sigma)!]\}$. Let $\mathcal{R}^k \subseteq X \times X$ be the binary relation such that for $x, x' \in X$, $(x, x') \in \mathcal{R}^k$. $x$ and $x'$are called control consistent w.r.t. $\Sigma_c^k$, if

$$E(x) \cap D^k(x') = E(x') \cap D^k(x) = \emptyset, \tag{14}$$
$$T(x) = T(x') \Rightarrow M(x) = M(x'). \tag{15}$$

The control cover $\mathcal{C}^k$, the induced generator $\mathbf{J^k}$, and the local controller $\mathbf{LOC^k}$ can be constructed, corresponding to the component $\mathbf{G^k}$**,** as proposed in [17].

In this paper, The disabled transitions set in **SUP** at an arbitrary state $x$is written as $D(x) = \bigcup_k D^k(x)$, where $k \in \mathcal{K}$ and $\mathcal{K}$ is Index set of local controllers and $D^k(x) = \{\sigma \in \Sigma_c^k \mid \neg\xi_k(x,\sigma)! \& (\exists s \in \Sigma^*)[\xi_k(x_0, s) = x \;\&\; \delta(q_0, s\sigma)!]\}$.Thus, (5) can be rewrite as follows,

$$E(x') \cap D(x) = E(x') \cap [\textstyle\bigcup_k D^k(x)] = \bigcup_k [E(x') \cap D^k(x)] = \emptyset. \tag{16}$$

Hence, $\forall k \in \mathcal{K}, E(x') \cap D^k(x) = \emptyset$. Since, the $k^{th}$ controllable event subset can only be disabled in the $k^{th}$ local controller, we can write $\forall j \neq k, E(x') \cap D^j(x) = E(x) \cap D^j(x') = \emptyset$, for each pair of states $(x, x')$ in the $k^{th}$ local controller. Thus, (5) is relaxed to $E(x') \cap D^k(x) = \emptyset$and $E(x) \cap D^k(x') = \emptyset$. We can write$\bigcup_{j\neq k} D^j(x) = D(x) \Longrightarrow E(x') \cap D^k(x) = \emptyset$ is true, even if $E(x') \cap D(x) \neq \emptyset$. It means that all controllable events which cannot be disabled by $k^{th}$ local controller, are self-looped at states where they are disabled by the monolithic supervisor. We call this new self-looped generator, as $\mathbf{S_k}, \forall k \in \mathcal{K}$. Each local controller, constructed in [17], can be obtained by supervisor reduction procedure computing on $\mathbf{S_k}, \forall k \in \mathcal{K}$. Moreover, this method is not restricted to localize a monolithic supervisor corresponding to each component of the plant. In fact, the proposed method can be carried out using any arbitrary partitioning of controllable events set. Moreover, components of the plant need not be defined on pairwise disjoint events sets. In a special case, this method can be used to localize a monolithic supervisor w.r.t. each controllable event of the plant. The number of states of each local controller, achieved for each controllable event of a component may be less than the state cardinality of the local controller corresponding to the component. Thus, this method may be more flexible than the supervisor localization procedure, proposed in [17], in terms of partitioning the controllable events set and in the state reduction point of view.

## 4. Construction of Distributed Supervisory Control Based on Coparanormality

Supervisor reduction/localization procedures are computed according to control consistency of states. In fact they are the methods based on state reduction. However, the main goal of this paper is to establish a relationship between observation properties of the monolithic supervisor and observation properties of local controllers.We define a new coobservation property, namely coparanormality, which is an extension of paranormality. Paranormality states that any subsequent unobservable event, whichoccursin the plant, does not exit the corresponding language. It is obvious that, if a specification $E$is $(L(\mathbf{G}), P) -$ paranormal, where $P: \Sigma^* \rightarrow \Sigma_0^*$ and $\Sigma_0 \subseteq \Sigma_c$, then $E$ is controllable w.r.t. $L(\mathbf{G})$ and $\Sigma_u$. Moreover, if only uncontrollable events are unobservable, then a controllable sublanguage $K_s \subseteq E$is $(L(\mathbf{G}), P) -$ paranormal.We define coparanormality as a decentralized control counterpart of paranormality property for supervisory control of discrete-event systems (DES) with partial observations. Paranormality is just controllability with respect to (w.r.t.) unobservable events. We introduce control authority of each local controller by coparanormality. Similar to paranormality and normality properties in the monolithic supervisory control, coparanormality is a counterpart of decomposability (not conormality) in distributed supervisory control.

*Definition 1*($Coparanormality$): The language $K$ is $(L(\mathbf{G}), P_k, \Sigma^k) -$ coparanormal, where $P_k: \Sigma^* \longrightarrow (\Sigma^k)^*, \forall k \in \mathcal{K}$ and $\mathcal{K}$ is index set,If $\forall k \in \mathcal{K}, \exists K_k, \overline{K}_k(\Sigma - \Sigma^k) \cap L(\mathbf{G}) \subseteq \overline{K}_k$ and $K = \bigcap_k K_k$.

Informally, coparanormality guarantees that any string in $\overline{K}$ followed by any unobservable events in $L(\mathbf{G})$, through several natural projections $P_k$, should remain in $\overline{K}$.

Recall paranormality of a supervisor [15]. If controllable events of a supervisor are observable, then the supervisor is paranormal under any projection channel. According to this statement, a set of local controllers can be constructed, each one observesa subset of controllable events. The supervisor $K_s$ can be distributed to a set of local controller $K_i$, where $P_i: \Sigma^* \longrightarrow (\Sigma^i)^*$, and $\overline{K}_i(\Sigma - \Sigma^i) \cap L(\mathbf{G}) \subseteq \overline{K}_i$, $\forall i \in I$ such that $\Sigma^i \subseteq (\bigcup_{j \neq i} \Sigma_c^j \cup \Sigma_u)$. The aim of this section is to distribute a monolithic supervisor $K_s$ to a set of local controllers $K_i$, $\forall i \in I$ and $I$ is the index set of local controllers, such that each $K_i$ can disable its corresponding controllable events only. Meanwhile, it is not guaranteed that each $K_i$ is a subset or equals to the global specification $E$. There is no restriction onpartitioning the controllable events set. Also, $\Sigma_c = \bigcup_i \Sigma_c^i$ and $\Sigma_c^i \cap \Sigma_c^j = \emptyset, i \neq j$. Now, we formulate distributed supervisory control based on coparanormality (DSCBC).

Let **SUP** be the recognizer of $K_s$, i.e. $K_s = L(\mathbf{SUP})$. In order to construct each local controller $K_i$, assume that $\mathbf{S_i} = (X, \Sigma, \xi_i, x_0, X_m)$ is a DES, constructed by editing the transition mapping of **SUP** as follows,

$$\xi_i(x,\sigma) := \begin{cases} \xi(x,\sigma), \text{if } \xi(x,\sigma)! \text{ or } \sigma \in \Sigma_c^i \\ x, \text{ if} \neg\xi(x,\sigma)!, \sigma \in \Sigma_c - \Sigma_c^i \text{and} (\exists s \in \Sigma^*), x = \xi(x_0,s), \delta(q_0,s\sigma)! \end{cases} \quad (17)$$

Informally, in order to construct $\mathbf{S_i}$, a transition which belongs to another subset of controllable events$\Sigma_c^j, j \neq i$ and is disabled at an arbitrary state of **SUP**, must become self-looped at this state. In other words, the control authority is restricted to one local controller. In fact, all controllable events except for corresponding controllable events, (i. e. $\forall \sigma \in \Sigma_c - \Sigma_c^i$) cannot be disabled at no state of the local controller. Note that, the partitioning of controllable events set $\Sigma_c$ is not restricted to event subsets of the plant components. Controllable events can be partitioned arbitrarily.In Proposition 1, we will prove that $K_s = \bigcap_i K_i$ and $\overline{K}_s = \bigcap_i \overline{K}_i$, where

$$K_i := L_m(\mathbf{S_i}) \cap L_m(\mathbf{G}), \quad (18)$$

$$\overline{K}_i := L(\mathbf{S_i}) \cap L(\mathbf{G}). \quad (19)$$

Since, all transitions in **SUP** are also in $\mathbf{S_i}$, and the set of marked states in $\mathbf{S_i}$ and **SUP** are identical, we can write $K_s \subseteq K_i, \forall i \in I$. Moreover, each local controller is paranormal, i.e. $\overline{K}_i(\Sigma - \Sigma^i) \cap L(\mathbf{G}) \subseteq \overline{K}_i, \forall i \in I$ where $P_i: \Sigma^* \longrightarrow (\Sigma^i)^*, \Sigma^i \subseteq (\bigcup_{j \neq i} \Sigma_c^j \cup \Sigma_u)$.

*Proposition 1:* Let $\mathbf{G}$ be a non-blocking plant, $K_s \neq \emptyset$ be a supervisor and $K_i, i \in I$be the local controller, which is constructed by (18) and (19). Then, $K_s = \bigcap_i K_i$ and $\overline{K}_s = \bigcap_i \overline{K}_i$.

*Proof:* We prove the claim in two parts.

Part 1: We should prove that (a) $K_s \subseteq \bigcap_i K_i$ and (b) $\overline{K}_s \subseteq \bigcap_i \overline{K}_i$

(a) Since$K_s \subseteq K_i$ , $\forall i \in I$, we can write $K_s \subseteq \bigcap_i K_i$.
(b) Since$K_s \subseteq \bigcap_i K_i$, we can write$\overline{K}_s \subseteq \overline{\bigcap_i K_i} \subseteq \bigcap_i \overline{K}_i$.

Part 2: We should prove that (a) $\overline{K}_s \supseteq \bigcap_i \overline{K}_i$ and (b) $K_s \supseteq \bigcap_i K_i$.

(a) Let $s \in \bigcap_i \overline{K}_i$.Then $s \in \overline{K}_i$. If $s = \varepsilon$as $\overline{K}_s \neq \emptyset$, then$s \in \overline{K}_s$. Suppose $s = \sigma_1$. Firstly, assume $\sigma_1 \in \Sigma_u$, then $s \in \overline{K}_s$, because $K_s$ is controllable. Now, assume$\sigma_1 \in \Sigma_c^i$. According to (17), if $\xi_i(x_0, \sigma_1)!$, then $\xi(x_0, \sigma_1)!$. If $\sigma_1 \in \Sigma_c - \Sigma_c^i$, then we can find $j \neq i, \sigma_1 \in \Sigma_c^j$. If$\xi_j(x_0, \sigma_1)!$, then $\xi(x_0, \sigma_1)!$. Thus, $s = \sigma_1$ does not cause an exit from $\overline{K}_s$. By repeating the foregoing argument we see that if $s = \sigma_1 \ldots \sigma_k \in \bigcap_i \overline{K}_i$ then $s \in \overline{K}_s$.
(b) Assume that$s \in \bigcap_i K_i$. From (18), $\xi_i(x_0, s) \in X_m, \forall i \in I$. According to part 2 (a), $\xi(x_0, s)!$. Therefore, $\xi(x_0, s) \in X_m$. It means that $s \in K_s$.

□

Apparently, this method does not reduce either state size, or event sizeof local controllers. We only stated that, a recognizer **SUP** can be distributed to a set of $\mathbf{S_i}, \forall i \in I$, and each $\mathbf{S_i}$ may be reduced by supervisor reduction procedure. It is a systematic procedure to construct a set of local controllers $\mathbf{LOC^i}$, which was constructed in [17], previously. The state size of each reduced state $\mathbf{S_i}$ (i.e. $\mathbf{LOC^i}$) may less than the state size of the reduced original supervisor. In this case, the original supervisor is called localizable, as stated in [17].

## 5. Relative Observability and Decomposability of a Supervisor

Observability and relative observability are properties of a language which imply that, decisions can be made consistently by observing look-alike strings through a projection channel. In the special case that a language is controllable, we have a (relative) observable supervisor.In the rest of the paper, we use coparanormality property to construct distributed supervisory control by self-looping the states in the feasible supervisor with controllable events, which are disabled by the supervisor except for a subset of events which can be disabled by the corresponding local controller.

Assume that $\mathbf{S_i^f} = (Y, \Sigma, \eta_i, y_0, Y_m)$ is a DES, constructed by editing the transition mapping of $\mathbf{SUP_f} = (Y, \Sigma, \eta, y_0, Y_m)$as follows,

$$\eta_i(y,\sigma) := \begin{cases} \eta(y,\sigma), \text{if } \eta(y,\sigma)! \text{ or } \sigma \in \Sigma_c^i \\ y, \text{if} \neg\eta(y,\sigma)!, \sigma \in \Sigma_c - \Sigma_c^i \end{cases} \quad (20)$$

Informally, (4) and (20) imply that, a transition$\sigma$ which is unobservable, or belongs to another subset of controllable events, and is disabled at an arbitrary state of $\mathbf{SUP_f}$, must be self-looped at that state. Meanwhile, such transition can be self-looped at other states $y$, where $(\exists s \in \Sigma^*), y = \eta(y_0, s), \neg\delta(q_0, s\sigma)!$.

In this section, we prove thatthe relative observability of a monolithic supervisoris a sufficient condition for its decomposition to several controllers; each one does not need to observe whole of the plant. In other words, a set of local controllers lead toa set of decomposed supervisor, in a top-down approach.

*Lemma 1:* Let $\mathbf{G}$ be a non-blocking plant and $\Sigma_0 \subseteq \Sigma$ be the observable event set. Let $\mathbf{SUP}$ be the recognizer of a supervisor $K_s$ and a set of local controllers $K_i$ be constructed by (18), (19). If $K_s$ is relative observable w.r.t. $(\bar{C}, \mathbf{G}, P)$ where $P: \Sigma^* \to \Sigma_0^*$, then $\forall \sigma \in \Sigma_c^i \cap (\Sigma - \Sigma_0)$ become self-loop transitions at all states of $\mathbf{S_j^f}, j \neq i$.

*Proof:* Assume$K_s$ is relative observable w.r.t. $(\bar{C}, \mathbf{G}, P)$. Assume $\mathbf{S_j^f}, j \neq i$ is constructed by (20) and $\sigma \in \Sigma_c^i \cap (\Sigma - \Sigma_0)$. For an arbitrary state $y$, we can write $\eta_i(y, \sigma) = y$, if $y = \eta_i(y_0, s)$& $[s\sigma \in \bar{K}_s$ or $s\sigma \in L(\mathbf{G}) - \bar{K}_s]$. Now, if $s\sigma \notin L(\mathbf{G})$, we can make $\sigma$ as self-looped transition at $y$. Therefore, $\sigma$ become self-loop transitions at all states of $\mathbf{S_j^f}, j \neq i$.

□

Lemma 1 declares that the controllable event set $\Sigma_c^i$ affects the behavior of the local controller $K_i$, only,i.e. it does not affect the behavior of other local controllers. If there exists $j \neq i$, such that $\exists \sigma' \in \Sigma_c^j \cap (\Sigma - \Sigma_0)$and$K_s$be relative observable w.r.t. $(\bar{C}, \mathbf{G}, P)$, then $K_s$ is decomposable w.r.t. $(L(\mathbf{G}), P_i, \Sigma_c^i), i = 1,2$ where $P_i: \Sigma^* \to (\Sigma^i)^*$,$\Sigma^1 = \Sigma - \{\sigma\}$ and $\Sigma^2 = \Sigma - \{\sigma'\}$.

*Theorem 1:* Let $\mathbf{G}$ be a non-blocking plant,along with observable event set $\Sigma_0 \subseteq \Sigma$. Let $K_s = L_m(\mathbf{SUP})$ be a relative observable supervisor w.r.t. $(\bar{C}, \mathbf{G}, P)$, where$P: \Sigma^* \to \Sigma_0^*$ and $\Sigma_c$ be partitioned into two local sets of controllable events $\Sigma_c^1$ and $\Sigma_c^2$. If $\exists \sigma, \sigma'$ s.t. $\sigma \in \Sigma_c^1 \cap (\Sigma - \Sigma_0), \sigma' \in \Sigma_c^2 \cap (\Sigma - \Sigma_0)$, then $K_s$ is decomposable w.r.t.$(L(\mathbf{G}), P_i, \Sigma_c^i), i = 1,2$, where $P_i: \Sigma^* \to (\Sigma^i)^*$,$\Sigma^1 = \Sigma - \{\sigma'\}$ and $\Sigma^2 = \Sigma - \{\sigma\}$.

*Proof:* Assume that$\sigma \in \Sigma_c^1 \cap (\Sigma - \Sigma_0), \sigma' \in \Sigma_c^2 \cap (\Sigma - \Sigma_0)$. According to Lemma 1, $\sigma'$ and $\sigma$ are self-looped at all states of $\mathbf{S_1^f}$ and $\mathbf{S_2^f}$, respectively. Thus, we can write

$$L_m(\mathbf{G}) \cap P_1^{-1} L_m(\mathbf{P_1}(\mathbf{S_1^f})) \cap P_2^{-1} L_m(\mathbf{P_2}(\mathbf{S_2^f})) = L_m(\mathbf{SUP}),$$
$$L(\mathbf{G}) \cap P_1^{-1} L(\mathbf{P_1}(\mathbf{S_1^f})) \cap P_2^{-1} L(\mathbf{P_2}(\mathbf{S_2^f})) = L(\mathbf{SUP}).$$

From (4) and (20), it is obvious $\mathbf{P_i}(\mathbf{S_i^f}) = \mathbf{P_i}(\mathbf{SUP}),\ i = 1,2$. Hence, we can write

$$L_m(\mathbf{G}) \cap P_1^{-1} L_m(\mathbf{P_1}(\mathbf{SUP})) \cap P_2^{-1} L_m(\mathbf{P_2}(\mathbf{SUP})) = L_m(\mathbf{SUP}),$$

$$L(\mathbf{G}) \cap P_1^{-1} L(\mathbf{P_1}(\mathbf{SUP})) \cap P_2^{-1} L(\mathbf{P_2}(\mathbf{SUP})) = L(\mathbf{SUP}).$$

Since $L_m(\mathbf{SUP}) \subseteq L_m(\mathbf{G}) \subseteq L(\mathbf{G})$ is true, we conclude that $L(\mathbf{G}) \cap P_1^{-1} L_m(\mathbf{P_1}(\mathbf{SUP})) \cap P_2^{-1} L_m(\mathbf{P_2}(\mathbf{SUP})) = L_m(\mathbf{SUP})$. It means that $K_s$ is decomposable w.r.t. $(L(\mathbf{G}), P_i, \Sigma_c^i)$, where $P_i: \Sigma^* \to (\Sigma^i)^*, i = 1,2$ and $\Sigma^1 = \Sigma - \{\sigma'\}$ and $\Sigma^2 = \Sigma - \{\sigma\}$.

□

Theorem 1 declares criteria to decompose a relative observable supervisor by appropriate partitioning the set ofcontrollable events. In such case, coparanormality of a supervisor leads to its decomposition.It means that we can define several natural projections, each one corresponding to a subset of controllable events, i.e. $P_i: \Sigma^* \to (\Sigma^i)^*, \Sigma^i = \Sigma - [\bigcup_{j \neq i} \Sigma_c^j \cap (\Sigma - \Sigma_0)]$, and $K_s = \bigcap_i P_i^{-1} P_i(K_s) \cap L(\mathbf{G})$.

In a special case, distribution of a supervisor can be carried out w.r.t. each component of the plant, such thatthe number of natural projections is equal to the number of components. In this case, decomposition of a monolithic supervisor means that each local controller corresponding to each component can make consistent decisions without observation of some controllable events corresponding to other components agents. This expresses the difference between coparanormality and decomposability of a supervisor.

*Corollary 1:*A relative observable supervisor is decomposable;if there are at least two controllable events, such that their observation by the supervisor does not affect the blocking.

## 6. Example- Localization and decomposition of the supervisory control ofguide way

Stations A and B are connected by a single one-way track from A to B, on a guide way as shown in Fig.1. The track consists of 4 sections, with stoplights (*) and detectors (!) installed at various section junctions [15]. Two vehicles $\mathbf{V_1}, \mathbf{V_2}$ use the guide way simultaneously. $\mathbf{V}_i, i = 1,2$ may be in state 0 (at A), state $j$ (while travelling in section $j = 1, \ldots, 4$), or state 5 (at B). The generator of $\mathbf{V}_i, i = 1,2$ are shown in Fig.2. In this example, the software package TCT [23] is used to modeling the plant and synthesize the supervisor.

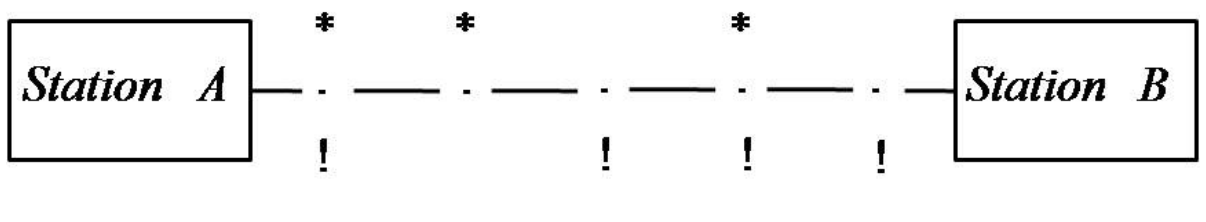

Fig. 1. Schematic of a guide way

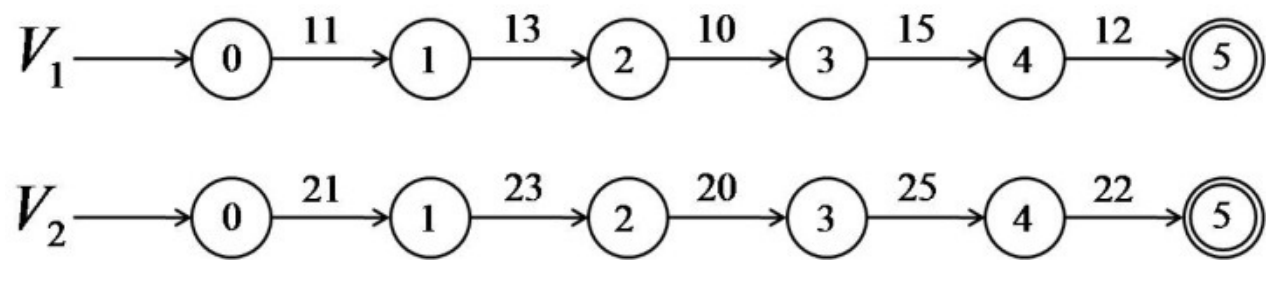

Fig. 2. DES model of each vehicle

The plant to be controlled is $\mathbf{G} = \mathbf{sync}(\mathbf{V_1}, \mathbf{V_2})$.In order to prevent collision, control of the stoplights must ensure that $\mathbf{V_1}$ and $\mathbf{V_2}$ never travel on the same section of track simultaneously. Namely, $\mathbf{V}_i$ , $i = 1,2$ are mutual exclusion of the state pairs $(i, i), i = 1, .., 4$. The supremal relative observable supervisor, where $P: \Sigma^* \to \Sigma_0^*$ , $\Sigma_0 = \Sigma - \{13,23\}$ and its reduced state formare shown in Figs. 3, 4, respectively. The control data is shown in Table 1. Moreover, the states and controllable events which are disabled/self-looped at states of each local controller are shown in Table 2.

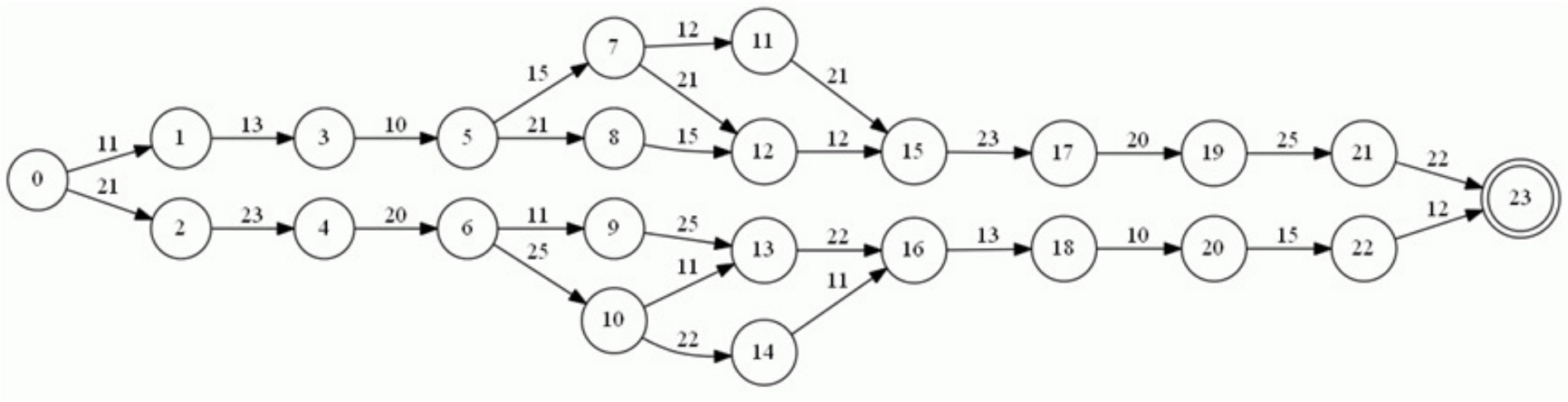

Fig. 3. The relative observable supervisor of guide way $(\mathbf{K_s})$

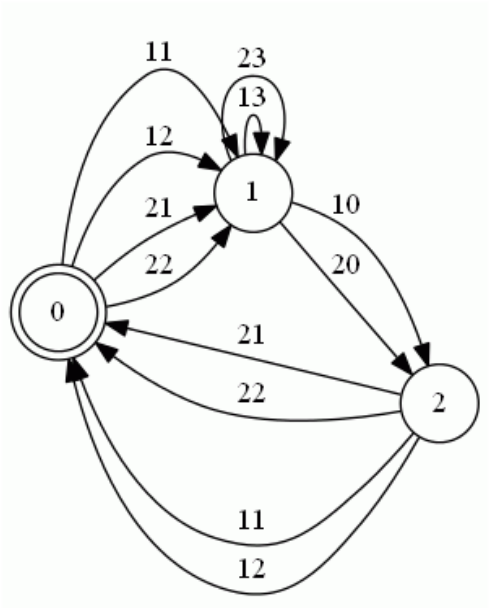

Fig. 4. The reduced supervisor of guide way

Table 1. Control data of the relative observable supervisor of guide way

| State | Disabled event | State | Disabled event |
|---|---|---|---|
| 1 | 21 | 2 | 11 |
| 3 | 21 | 4 | 11 |
| 8 | 23 | 9 | 13 |
| 12 | 23 | 13 | 13 |

Table 2. The states and controllable events which are disabled/self-looped at states of each local controller

(a) $\mathrm{V_1}$

| State | Disabled event | State | Self-looped event |
|---|---|---|---|
| 2 | 11 | 1 | 21 |
| 4 | 11 | 3 | 21 |
| 9 | 13 | 8 | 23 |
| 13 | 13 | 12 | 23 |

(b) $\mathrm{V_2}$

| State | Disabled event | State | Self-looped event |
|---|---|---|---|
| 1 | 21 | 2 | 11 |
| 3 | 21 | 4 | 11 |
| 8 | 23 | 9 | 13 |
| 12 | 23 | 13 | 13 |

The local controllers of $\mathrm{V_1}$ and $\mathrm{V_2}$ are shown in Figs. 5, 6, respectively. Each local controller can be reduced by the supervisor reduction procedure. The reduced local controllers are shown in Fig. 7. Clearly, the number of states in each local controller is reduced from 24 states to 3 states. The reduced local controllers which are shown in Fig. 7 are equal to the local controllers which can be constructed by supervisor localization procedure, proposed in [15].

Moreover, we show that $\mathbf{K_s}$ can be self-looped with all disabled events at the states, where they are disabled by $\mathbf{K_s}$ except for one controllable event which is not disabled by the corresponding local controller. The control data is shown corresponding to each controllable event in Table 3. The local controller corresponding to event 11 is shown in

Fig. 8. Events 13, 21 and 23 are self-looped at the states, where they were disabled by the monolithic supervisor.

The reduced state local controller corresponding to event 11 is constructed by supervisor reduction procedure. The automaton of this controller is shown in Fig. 8. Other local controllers are not shown in this figure. But the reduced ones are shown in Figs. 9 (b)-(d). Since events 15, 25 are not disabled by the monolithic supervisor, and each pair of states, where those events occur in between are control consistent, the reduced state local controller corresponding to those events are shown as one state automaton in Fig. 9 (e).

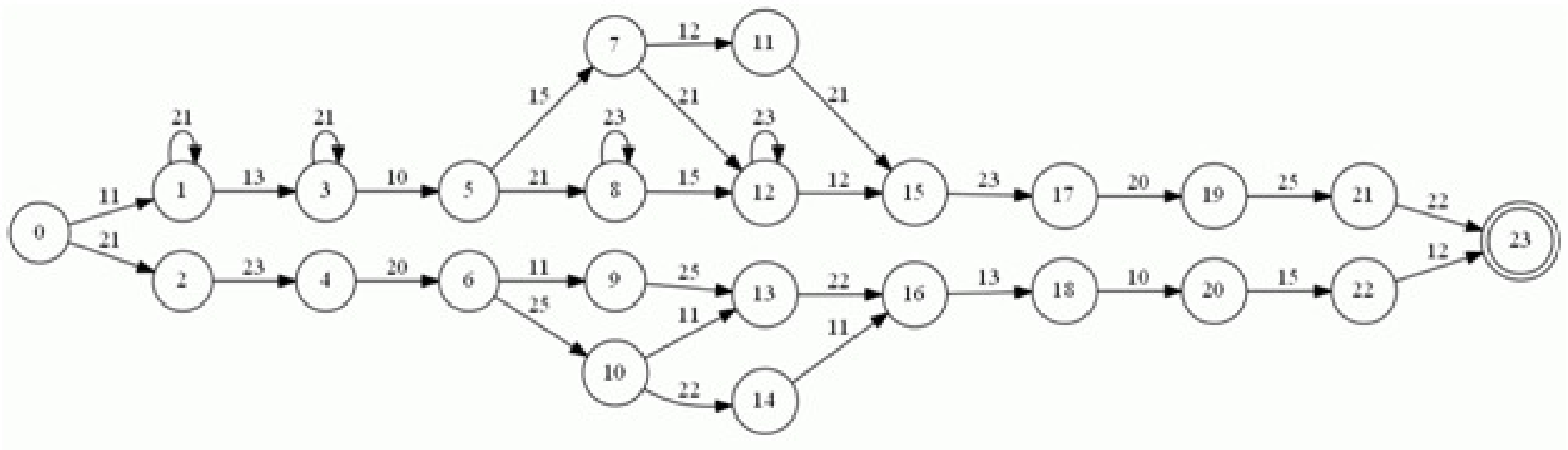

Fig. 5. Local controller of $V_1$

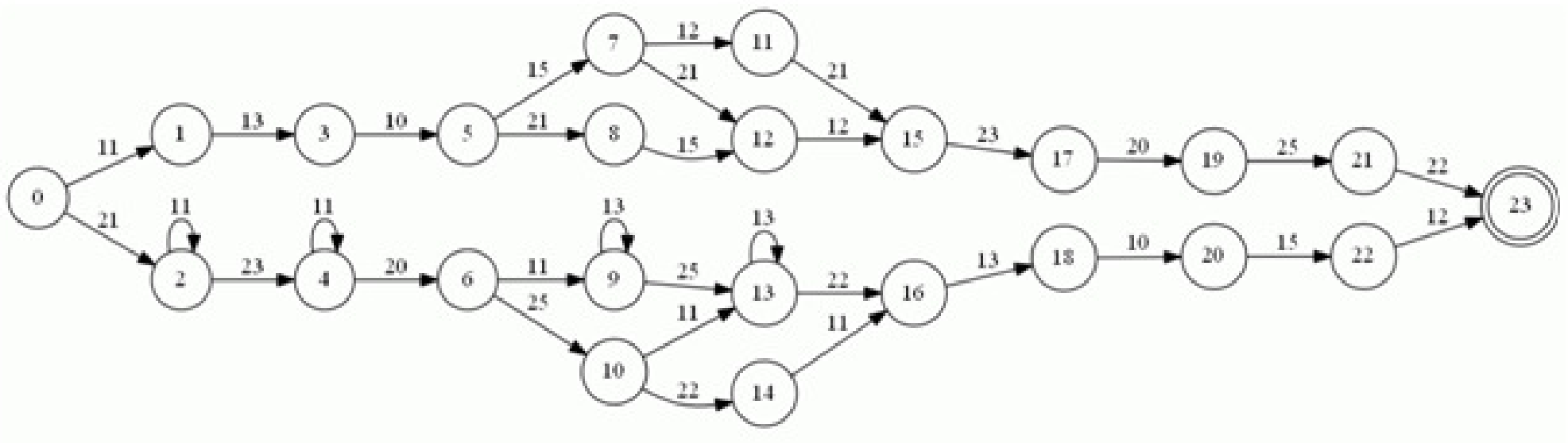

Fig. 6. Local controller of $V_2$

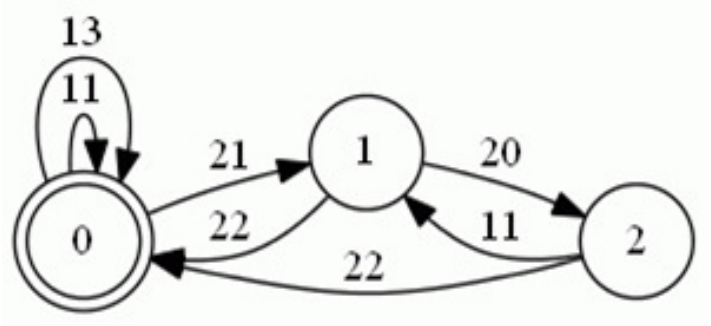

(a) The reduced local controller of $V_1$

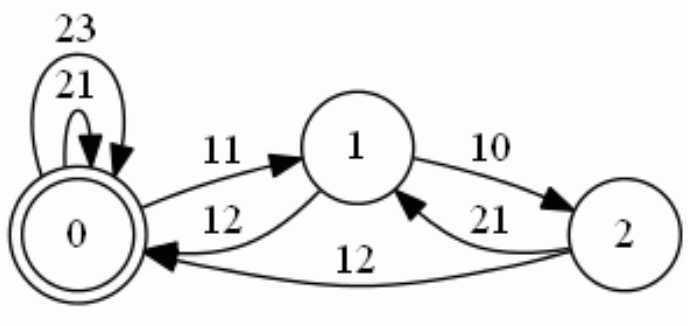

(b) The reduced local controller of $V_2$

Fig. 7. The reduced local controllers of components of guide way

Table 5. The states and controllable events which are disabled by local controllers

(a) Event 11

| State | Disabled event |
|---|---|
| 2 | 11 |
| 4 | 11 |

(b) Event 13

| State | Disabled event |
|---|---|
| 9 | 13 |
| 13 | 13 |

(c) Event 21

| State | Disabled event |
|---|---|
| 1 | 21 |
| 3 | 21 |

(d) Event 23

| State | Disabled event |
|---|---|
| 8 | 23 |
| 12 | 23 |

(e) Events 15, 25

| |
|---|
| Events 15, 25 is disabled at no state |

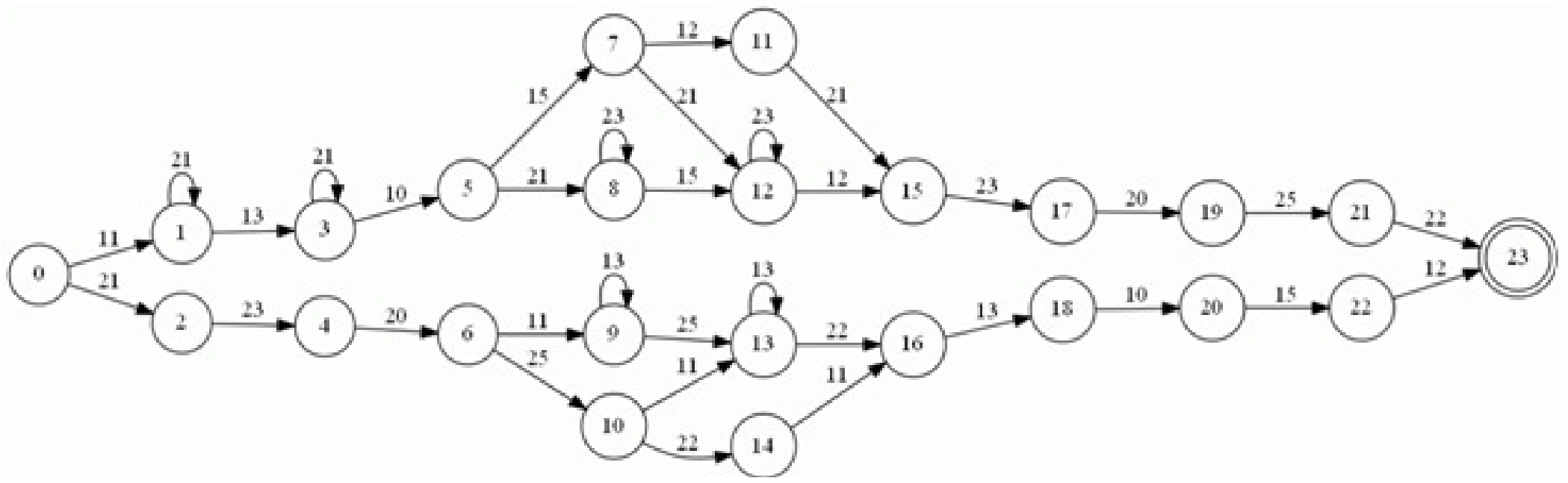

Fig. 8. Local controllers of guide way, corresponding to event 11

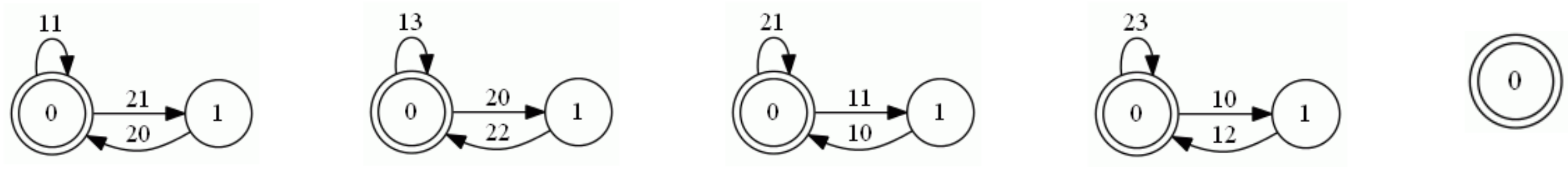

(a) (b) (c) (d) (e)

Fig. 9. Local controllers corresponding to each controllable events of guide way

(a) event 11 (b) event 13 (c) event 21 (d) event 23 (e) events 15, 25

Furthermore, the monolithic supervisor $K_s \coloneqq L(\mathbf{K_s})$ is decomposable with $P_1\colon \Sigma^* \to \Sigma_1^*$, $\Sigma_1 = \Sigma - \{13\}$ and $P_2\colon \Sigma^* \to \Sigma_2^*$, $\Sigma_2 = \Sigma - \{23\}$. They are shown in Figs. 10, 11.

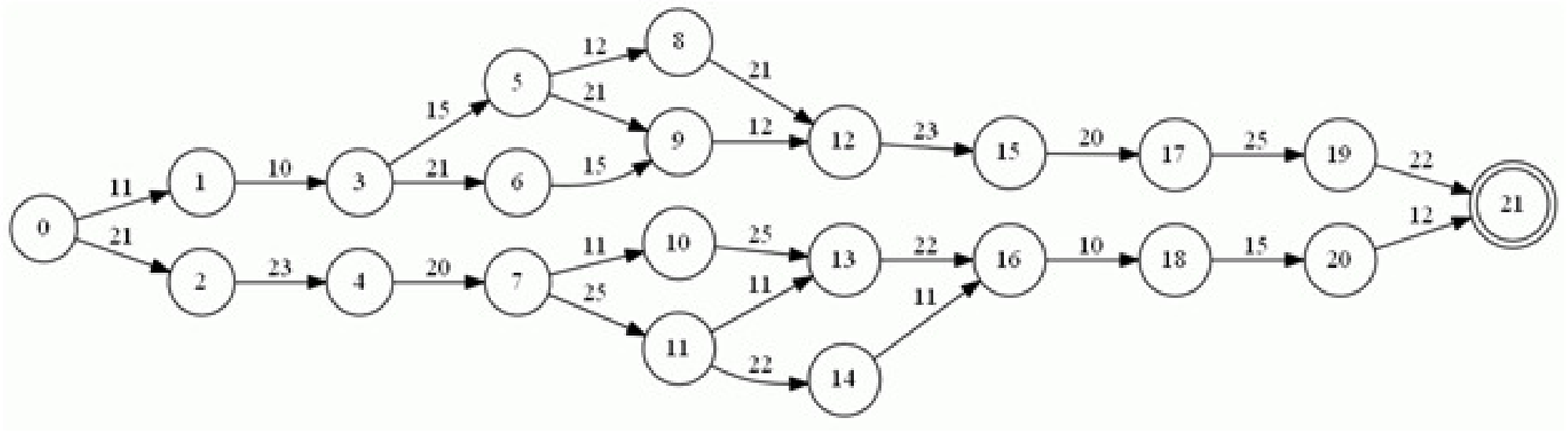

Fig. 10. Decentralized supervisor of guide way, $\mathbf{P_1}(\mathbf{K_s})$

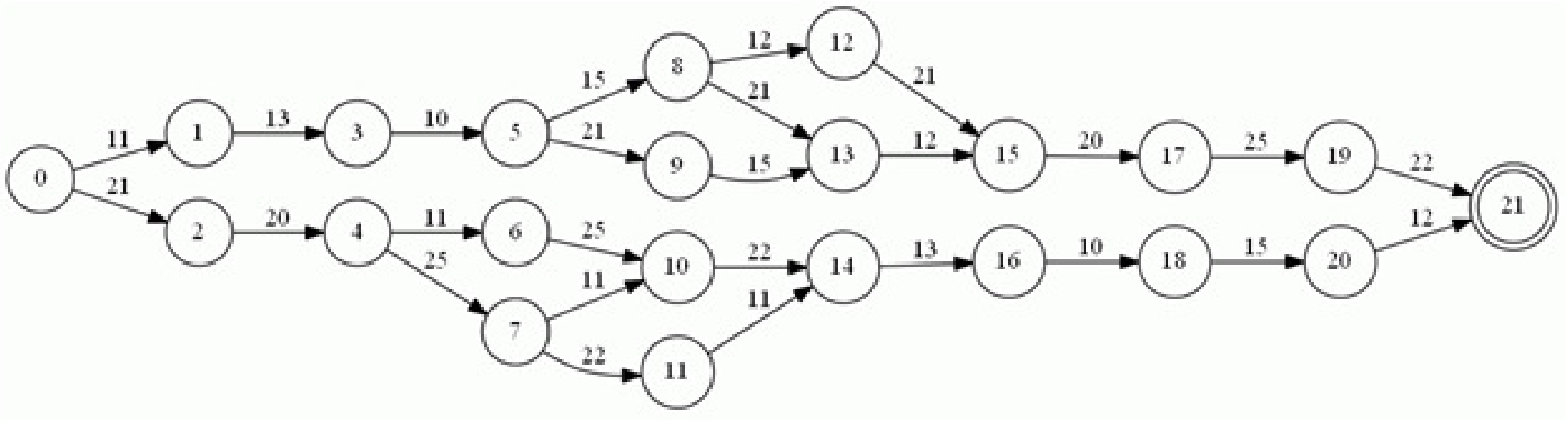

Fig. 11. Decentralized supervisor of guide way, $\mathbf{P_2}(\mathbf{K_s})$

## 7. Conclusions

Thispaper addresses a method to distribute amonolithic supervisor based on a new observation property, namely coparanormality. However, the control authority of local controllers is strictly local; unlike the decomposition, coparanormality imposes no restriction on observation scope oflocal controllers. It was shown that each supervisor can be coparanormal with proper definition of natural projections. It was proved that, relative observability property is a sufficient condition for decomposability of a supervisor. We showed that each reduced local controller, computed by the proposed method in this paper, is control equivalent tothe one, computed by supervisor localization procedure [17], w.r.t. the plant. Moreover, we generalized the supervisor localization procedure to find a set of local controllers corresponding to any arbitrary partition of controllable events set. In the special case each partition may have one controllable event. This method can be beneficial, when each component has more than one controllable event. Also, the components of the plant need not to have disjointed events. The implementation of such local controllers on programmable logic controllers (PLC) would become easier in industrial applications.